\documentclass[prb,twocolumn,amsmath,amssymb,showpacs]{revtex4}

\usepackage{bm}

\usepackage{longtable}
\usepackage{hyperref}
\usepackage[dvips]{graphicx}

\newcommand{\dummy}{\rule[-1ex]{0pt}{4ex}}

\begin{document} 

\title{{\it Ab initio} calculations of third-order elastic constants 
         and related properties for selected semiconductors} 

\author{Micha{\l} {\L}opuszy{\'n}ski}
\affiliation{Interdisciplinary Centre for Mathematical and Computational
             Modelling, University of Warsaw, Pawi{\'n}skiego 5A, 02-106 Warsaw,
             Poland }

\author{Jacek A. Majewski}
\affiliation{Institute of Theoretical Physics, Faculty of Physics,
             University of Warsaw, Ho{\.z}a 69, 00-681 Warsaw, Poland}


\date{\today}

\begin{abstract}
We present theoretical studies for the third-order elastic constants $C_{ijk}$
in zinc-blende nitrides AlN, GaN, and InN. Our predictions for these compounds
are based on detailed {\it ab initio} calculations of strain-energy and
strain-stress relations in the framework of the density functional theory. To
judge the computational accuracy, we compare the {\it ab initio} calculated
results for $C_{ijk}$ with experimental data available for Si and GaAs. We also
underline the relation of the third-order elastic constants to other quantities
characterizing  anharmonic behaviour of materials, such as pressure derivatives
of the second-order elastic constants $c'_{ij}$ and the mode Gr\"uneisen
constants  for long-wavelength acoustic modes $\gamma({\bf q}, {\bf j})$. 
\vspace{1.5ex}
\newline
\emph{Paper accepted to Phys. Rev. B.}
\end{abstract} 

\maketitle

\section{Introduction}
Third-order elastic constants $C_{ijk}$ are important quantities characterizing
nonlinear elastic properties of materials and the interest in them dates back
to the beginning of modern solid state physics.  \cite{Birch1947,
Murnaghan1951, Bhagavantam1966, Thurston1964, Brugger1964} Third- and
higher-order elastic constants are useful not only in describing mechanical
phenomena when large stresses and strains are involved (e.g.,  in
heterostructures of optoelectronic devices), but they can also serve as a basis
for discussion of other anharmonic properties. The applications include
phenomena such as thermal expansion, temperature dependence of elastic
properties, phonon-phonon interactions etc.  \cite{Hiki1981} 

As far as theoretical studies are concerned, at the beginning the third-order
elastic constants were calculated in the framework of the valence force Keating
model. \cite{Keating1966a}  Later on, many other more sophisticated microscopic
theories were employed to describe and predict nonlinear elastic properties of
crystals on the basis of their atomic composition.\cite{Hiki1981}  Nowadays,
precise {\it ab initio} calculations seem to be the most promising approach to
handle this task. Such applications of density functional theory (DFT) on 
the local density approximation level (LDA) were already reported. 
\cite{Nielsen1985,Sorgel1998}  

Recently, one observes increased interest in nonlinear effects in elastic
\cite{Kato1994,Lepkowski2004,Lepkowski2005} and piezoelectric properties.
\cite{Shimada1998, Bester2006} This is strongly connected to the fact that
research focuses nowadays on the semiconductor nanostructures.  In such systems
these nonlinear effects are not only more pronounced than in bulk materials,
but very often their reliable quantitative description  is a prerequisite for
correct theoretical explanation of the experimental data.
\cite{Frogley2000,Ellaway2002, Ma2004, Luo2005,Lepkowski2005,Bester2006}  In
this paper, we perform {\it ab initio} calculations of the unknown third-order
elastic constants in cubic nitrides.  The nitrides are technologically
important group of materials for which the nonlinear effects are particularly
significant. \cite{Kato1994,Vaschenko2003, Lepkowski2005, Lepkowski2006}
Therefore, the knowledge of the third-order elastic moduli will definitely
improve the modeling of nitride based nanostructures. In this work we also
briefly discuss the applications of $C_{ijk}$ to  determination of other
anharmonic properties, namely, pressure derivatives of second-order elastic
moduli $c'_{ij}$ and mode Gr\"uneisen constants $\gamma({\bf q},{\bf j})$.
Since the third-order effects are rather subtle, their computational
determination can also serve as a precise test of accuracy for modern {\it ab
initio} codes based on DFT approach.
 
The paper is organized as follows. In Sec. \ref{Overview} we give a general
overview of the nonlinear elasticity theory.  Sec. \ref{ElasticConstants}
contains a description of employed methodology. Also results for third-order
elastic constants obtained from {\it ab initio} calculations are presented
there.  Our findings for Si and GaAs are compared with previous numerical
calculations and measurements, later on theoretical predictions for zinc-blende
nitrides AlN, GaN, and InN are given.  Secs. \ref{PressureConstants} and
\ref{GrueneisenConstants} deal with the determination of quantities related to
third-order elastic constants, namely, the pressure dependent elastic constants
and mode Gr\"uneisen constants, respectively. Finally, we conclude the paper in
Sec. \ref{Conclusions}. 

\section{Overview of nonlinear elasticity theory \label{Overview}}
Here we will recall some basic facts from nonlinear theory of elasticity.
\cite{Birch1947, Murnaghan1951, Bhagavantam1966, Thurston1964, Brugger1964,
Hiki1981}  Let us consider point ${\bf a}$ which, after applying strain to a
crystal, moves to the position ${\bf x}$. After introducing the Jacobian
matrix ${\bf J}$
\begin{equation}
    J_{ij}=\frac{\partial x_i}{\partial a_j} 
\end{equation}
we may define the Lagrangian strain 
\begin{equation}
    \label{definitionEta}
    \bm{\eta} = \frac{1}{2} ( \bm{J}^T \bm{J} - {\bf 1} ),
\end{equation}
which is a convenient measure of deformation for an elastic body.

The energy per unit mass $E(\bm{\eta})$ corresponding to the applied strain 
may be developed in power series with respect to $\bm{\eta}$. This leads to 
the expression
\begin{equation}
\label{energyStrainRel1}
\rho_0 E(\bm{\eta}) =
   \frac{1}{2!}\sum_{i,j=1,6}  c_{ij} \eta_i \eta_j +
          \frac{1}{3!}\sum_{i,j,k=1,6} C_{ijk}  \eta_i \eta_j \eta_k+
          \dots,
\end{equation}
where we applied Voigt convention ($\eta_{11} \to \eta_1$, $\eta_{22} \to
\eta_2$, $\eta_{33} \to \eta_3$, $\eta_{23} \to \eta_4/2$,  $\eta_{13} \to
\eta_5/2$, $\eta_{12} \to \eta_6/2$) and introduced the density of unstrained
crystal $\rho_0$.  The $c_{ij}$ and $C_{ijk}$ denote here second- and
third-order elastic constants respectively.        
\footnote{
         In older texts concerning nonlinear elasticity
         different definitions of $C_{ijk}$ may be encountered.
         In this paper we follow the convention proposed in 
         Ref. \onlinecite{Brugger1964} which is now a standard approach.
        }   
If we introduce $\bm{J}=( \bm{1} + \bm{\epsilon})$ and assume that
$\bm{\epsilon}$ is symmetric (rotation free) linear strain tensor, the
definition of $\eta$ [Eq. (\ref{definitionEta})]  yields
\begin{equation}
    \label{expansionEta}
    \bm{\eta}= \bm{\epsilon} + \frac{1}{2}\bm{\epsilon}^2.
\end{equation}
Substituting the above result to the expansion in Eq. (\ref{energyStrainRel1}) 
and leaving only terms up to second order with respect to components of
$\bm{\epsilon}$ recover the infinitesimal theory of elasticity. 

Naturally, the general expression for energy of strained crystal, as given by 
Eq. (\ref{energyStrainRel1}), can be simplified by employing symmetry
considerations.  For cubic crystals, this procedure yields the following
formula:
{ \small
\begin{eqnarray}
   \label{energyStrainRel2}
   \rho_0 E(\bm{\eta}) 
&=&\frac{1}{2} c_{11} \left(\eta_1^2+\eta_2^2+\eta_3^2\right)+\frac{1}{2} 
    c_{44} \left(\eta_4^2+\eta_5^2+\eta_6^2\right) + \nonumber  \\ 
& &+c_{12} (\eta_1 \eta_2+\eta_3 \eta_2+\eta_1 \eta_3) + \nonumber \\
& & \frac{1}{6} C_{111} \left(\eta_1^3+\eta_2^3+\eta_3^3\right) +  \\
& & \frac{1}{2} C_{112} \left(\eta_2 \eta_1^2+\eta_3 \eta_1^2+\eta_2^2 
    \eta_1+\eta_3^2 \eta_1+\eta_2 \eta_3^2+\eta_2^2 \eta_3\right) + \nonumber \\
& &  C_{123} \eta_1 \eta_2 \eta_3 + \frac{1}{2} C_{144} \left(\eta_1 
    \eta_4^2+\eta_2 \eta_5^2+\eta_3 \eta_6^2\right)+ \nonumber \\
& & \frac{1}{2} C_{155} \left(\eta_2 \eta_4^2+\eta_3 \eta_4^2+\eta_1 
    \eta_5^2+\eta_3 \eta_5^2+\eta_1 \eta_6^2+\eta_2 \eta_6^2\right)+ 
    \nonumber \\
& & C_{456} \eta_4 \eta_5 \eta_6 + \dots \nonumber
\end{eqnarray}
}

Another fundamental quantity in the theory of finite deformations is 
Lagrangian stress
\begin{equation}
 \label{stressStrainRel1}
  t_{ij}= \rho_0 \frac{\partial E}{\partial \eta_{ij}},
\end{equation}
which can be expressed in terms of linear stress tensor $\bm{\sigma}$ using
the following formula
\begin{equation}
 \label{stressStrainRel2}
 \bm{t}= \det(\bm{J}) \bm{J}^{-1} \bm{\sigma} \left ( \bm{J}^T \right )^{-1}.
\end{equation}
Again, Voigt convention ($t_{11} \to t_1$, $t_{22} \to t_2$, $t_{33} \to t_3$,
$t_{23} \to t_4$, $t_{13} \to t_5$, $t_{12} \to t_6$) is used here. 

\section{ Determination of third-order elastic constants 
          \label{ElasticConstants} }

\subsection{Methodology and computational details}
In this work, we have determined third-order elastic constants for Si, GaAs,
and zinc-blende nitrides (AlN, GaN, and InN) on the basis of quantum DFT
calculations  for deformed crystals.  The results were obtained in two ways -
employing strain-energy formula [Eq. (\ref{energyStrainRel2})] and from 
strain-stress relation [Eqs. (\ref{stressStrainRel1}) and
(\ref{stressStrainRel2})].

The detailed procedure was as follows. We considered  six sets of deformations
parametrized by $\eta$ 
\begin{eqnarray}
\bm{\eta}_A&=&(\eta,0,0,0,0,0),       \nonumber \\
\bm{\eta}_B&=&(\eta,\eta,0,0,0,0),    \nonumber \\
\bm{\eta}_C&=&(\eta,0,0,\eta,0,0),    \nonumber \\
\bm{\eta}_D&=&(\eta,0,0,0,\eta,0),              \\
\bm{\eta}_E&=&(\eta,\eta,\eta,0,0,0), \nonumber \\
\bm{\eta}_F&=&(0,0,0,\eta,\eta,\eta). \nonumber 
\end{eqnarray}
In every case, $\eta$ was varied between $-0.08$ and $0.08$ with step $0.008$.
For every deformed configuration, the positions of atoms were optimized and both
energy and stress tensors were calculated on the basis of  quantum DFT
formalism.  In this way, for each type of distortion, dependencies of energy
$E(\eta)$ and stress tensor $\bm{t}(\eta)$ on strain parameter $\eta$ were
obtained. The numerical results have been in turn compared with the expressions
from the nonlinear theory of elasticity, which are summarized in Table
\ref{tableFormulas}. This allows to extract the values of the second- and 
third-order elastic constants, by performing suitable polynomial fits.

\begin{table*}[!ht]
\caption{Dependencies of energy and stress on deformation parameter $\eta$  
         for considered types of deformation $\bm{\eta}_A,\dots, \bm{\eta}_F$,
         which have been used to determine second- and third-order
         elastic constants.}
         \label{tableFormulas}
\begin{center}
\begin{tabular}{ll}
Energy:  &  \\
        & $\rho_0 E(\bm{\eta}_A) = 
            \frac{1}{6} C_{111}\eta^3 +\frac{1}{2} c_{11} \eta^2 
                        \doteq f_A(\eta) $ \\
  
        & $ \rho_0 E(\bm{\eta}_B) =
            \left(\frac{1}{3}C_{111}+C_{112}\right) \eta^3+
            (c_{11}+c_{12}) \eta^2 \doteq f_B(\eta)$\\
  
        & $ \rho_0 E(\bm{\eta}_C) =
            \left(\frac{1}{6} C_{111} + \frac{1}{2} C_{144} \right) \eta^3+
            \left(\frac{1}{2} c_{11} + \frac{1}{2} c_{44} \right) \eta^2 
                        \doteq f_C(\eta) $\\

        & $ \rho_0 E(\bm{\eta}_D) = 
            \left(\frac{1}{6} C_{111} + \frac{1}{2} C_{155} \right) \eta^3+
            \left(\frac{1}{2} c_{11} + \frac{1}{2} c_{44} \right) \eta^2 
                        \doteq f_D(\eta)$ \\
  
        & $ \rho_0 E(\bm{\eta}_E) =
            \left(\frac{1}{2} C_{111} + 3C_{112} + C_{123}\right) \eta^3+
            \left( \frac{3}{2} c_{11} + 3 c_{12} \right) \eta^2 
                        \doteq f_E(\eta)$ \\
  
        & $ \rho_0 E(\bm{\eta}_F) =
            C_{456} \eta^3+\frac{3}{2} c_{44} \eta^2 
                        \doteq f_F(\eta)$  \\
Stress:  & \\
        & $ t_{1}( \bm{\eta}_A ) =
            \frac{1}{2} C_{111} \eta^2 + c_{11} \eta \doteq g_{A1}(\eta) $ \\

        & $ t_{2}( \bm{\eta}_A ) =
            \frac{1}{2} C_{112} \eta^2 + c_{12} \eta \doteq g_{A2}(\eta)$ \\
        
        & $ t_{3}( \bm{\eta}_B ) = 
          \left ( C_{123} + C_{112} \right ) \eta^2 + 2 c_{12} \eta 
                  \doteq g_B(\eta)$ \\
               
        & $ t_{4}( \bm{\eta}_C ) = 
            C_{144} \eta^2 + c_{44} \eta \doteq g_C(\eta) $ \\
                                
        & $ t_{5}( \bm{\eta}_D ) = 
                C_{155} \eta^2 + c_{44} \eta \doteq g_D(\eta)$ \\
                                
        & $ t_{4}( \bm{\eta}_F ) = 
                C_{456} \eta^2 + c_{44} \eta \doteq g_F(\eta)$
\end{tabular}
\end{center}
\end{table*}

The DFT calculations have been performed using the {\it ab initio} total energy
code VASP developed at the Institut f\"ur Materialphysik of Universit\"at Wien.
\cite{Kresse1993,Kresse1996,Kresse1996a} The projector augmented wave (PAW)
approach \cite{Blochl1994} has been used in its variant available in the VASP
package. \cite{Kresse1999} For the exchange-correlation functional generalized
gradient approximation (GGA) according to Perdew, Burke and Ernzerhof (PBE)
\cite{Perdew1996, Perdew1996a} has been applied. For Ga and In, semicore 3d 
and 4d electrons have been explicitly included in the calculations. 

\begin{figure}[!ht]
  \includegraphics[height=0.45\textwidth,angle=270]
        {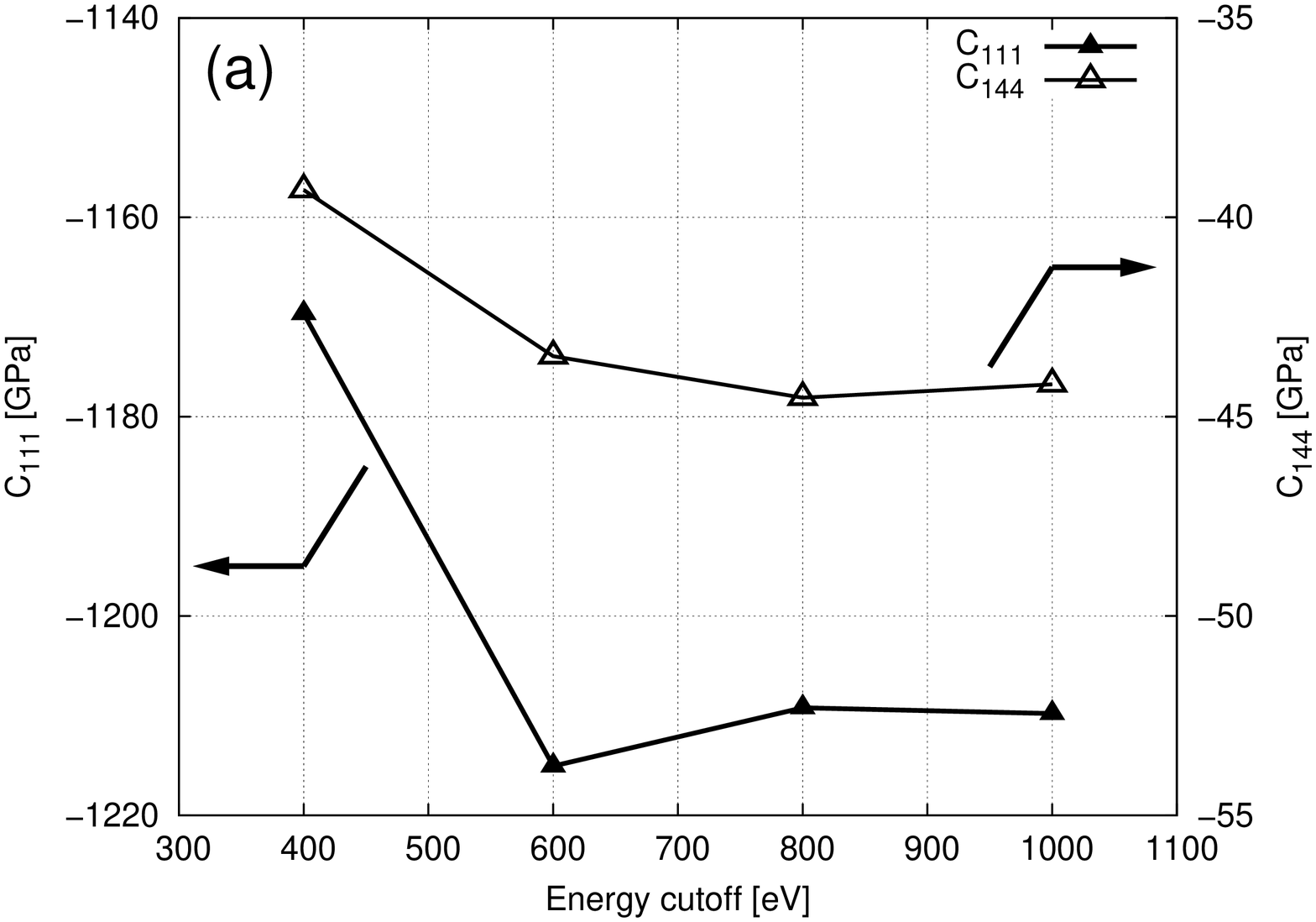}
  \includegraphics[height=0.45\textwidth,angle=270]
        {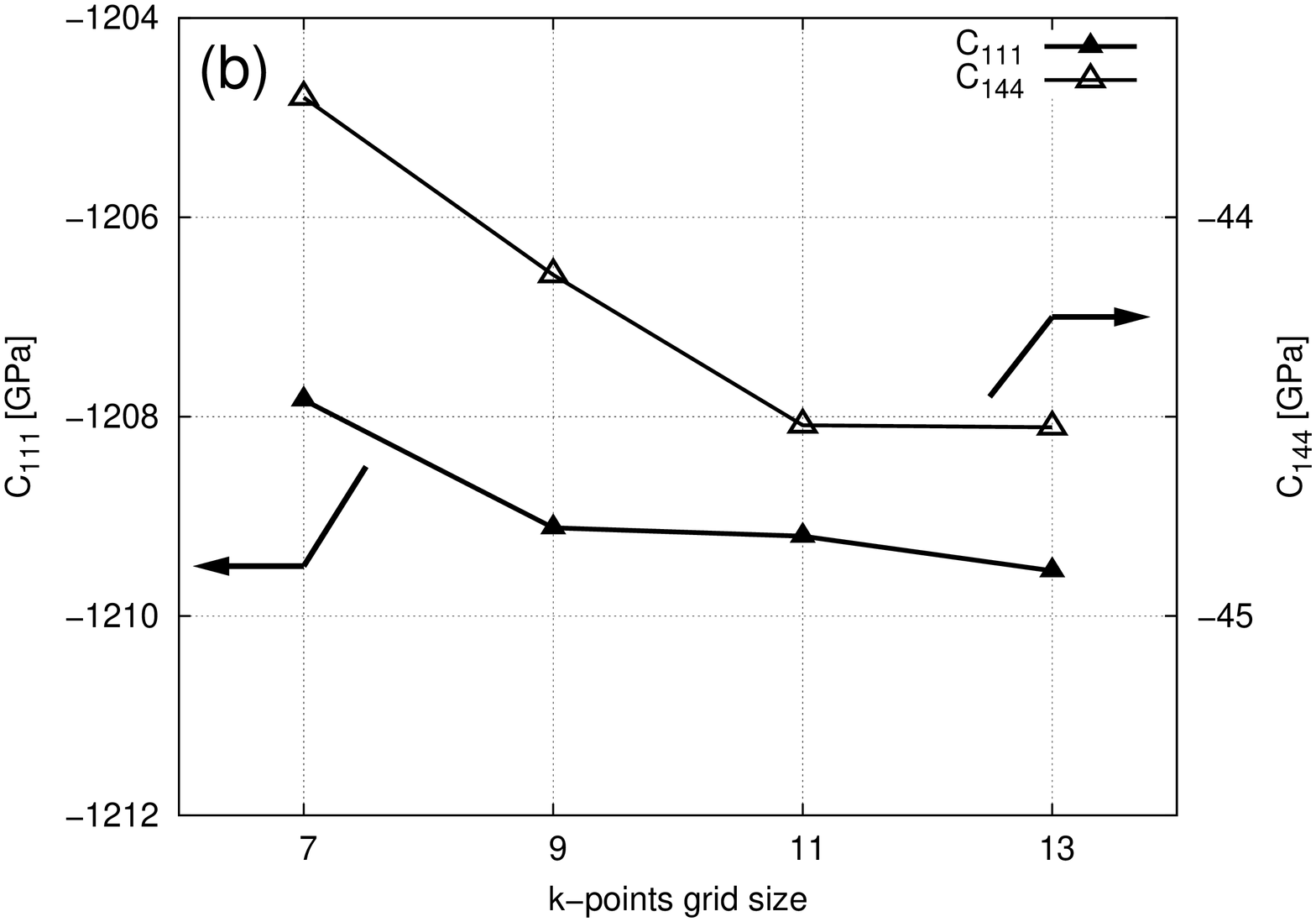}
    \caption{Sample convergence tests for the third-order elastic
			 constants in zinc-blende GaN.  Panel (a) illustrates the
			 dependence of $C_{111}$ and $C_{144}$ on energy cutoff
			 (Monkhorst-Pack sampling $11 \times 11 \times 11$ was applied for
			 all points).  Panel (b) shows the analogous dependence on the
			 density of k-points mesh (energy cutoff 800 eV was used for all
			 points).  Note different scales for $C_{111}$ and $C_{144}$.
			 Later on, all calculations for GaN have been performed with the
			 energy cutoff 800 eV and the Monkhorst-Pack sampling
			 $11\times11\times11$. 
             \label{figureConvergence}        
    }
\end{figure}

Since the determination of subtle third-order effects requires high precision,
we have performed careful convergence tests for parameters governing the
accuracy of computations.  On the basis of our tests we have chosen the
following energy cutoffs $E_{\rm{cutoff}}^{\rm{Si}} = 600\;\rm{eV}$,
$E_{\rm{cutoff}}^{\rm{GaAs}} = 700\;\rm{eV}$, and  $E_{\rm{cutoff}}^{\rm{AlN}}
= E_{\rm{cutoff}}^{\rm{GaN}} = E_{\rm{cutoff}}^{\rm{InN}} = 800\;\rm{eV}$.  For
the Brillouin zone integrals we have followed the Monkhorst-Pack
scheme\cite{Monkhorst1976}, in Si and GaAs we have used $13\times13\times13$
mesh, whereas for AlN, GaN, and InN we have applied $11\times11\times11$
sampling. One example of performed tests for GaN is presented in Fig.
\ref{figureConvergence}. It illustrates the dependence of two sample elastic
moduli $C_{111}$ and $C_{144}$ on the energy cutoff and density of
Monkhorst-Pack k-point mesh. For the chosen parameters
($E_{\rm{cutoff}}^{\rm{GaN}}=800\;\rm{eV}$ and $11\times11\times11$ k-point
mesh) the difference between successive values of examined constants in our
test is lower than $1\;\rm{GPa}$.  This difference is smaller than e.g.
discrepancies observed between results obtained from strain-energy and
strain-stress approach which, in the opinion of the authors, indicates that the
convergence with respect to parameters responsible for numerical accuracy is
very reasonable. 

\subsection{Results and discussion}
\begin{figure}[!ht]

  \includegraphics[height=0.45\textwidth,angle=270]
        {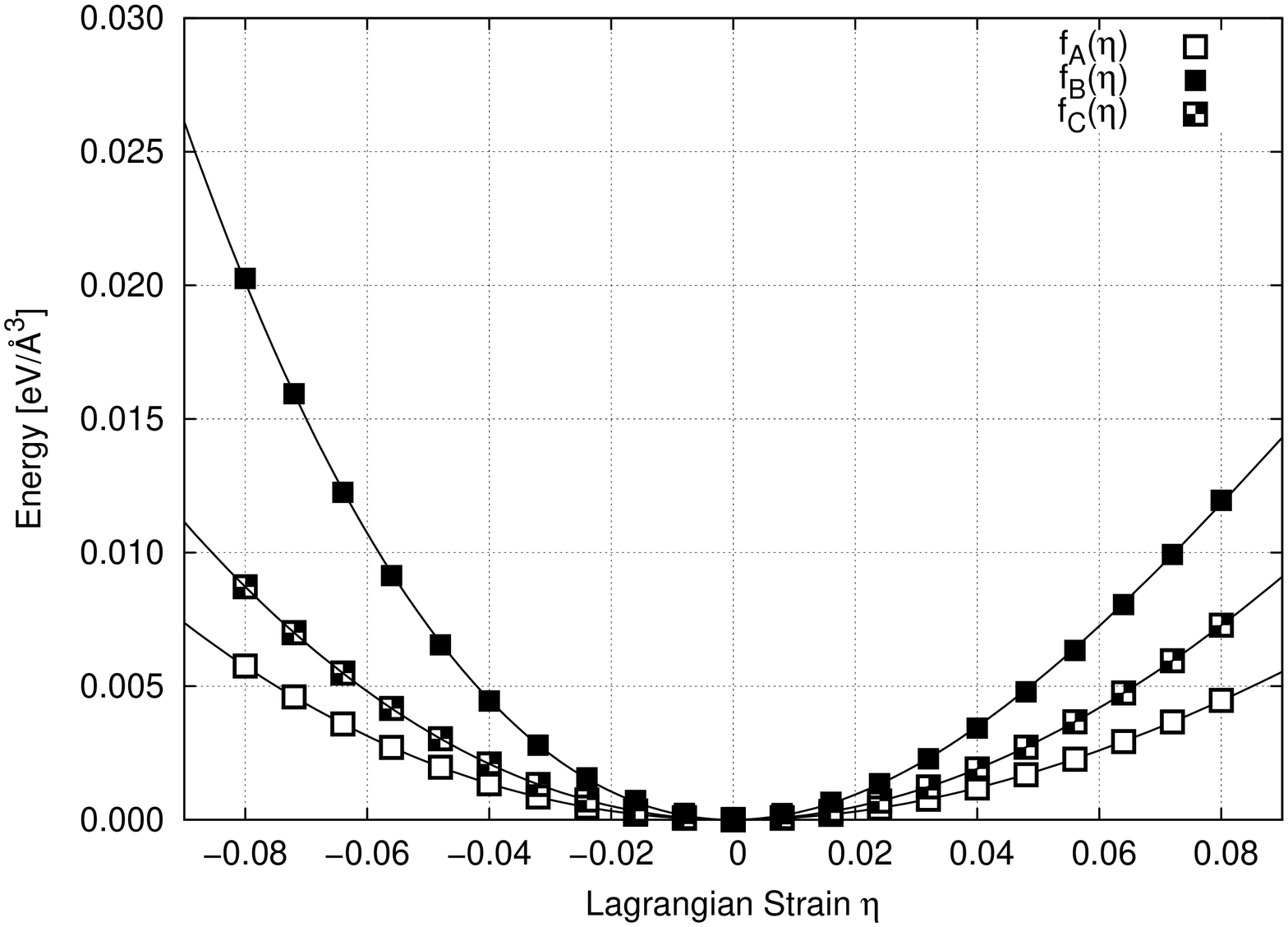}
  \includegraphics[height=0.45\textwidth,angle=270]
       {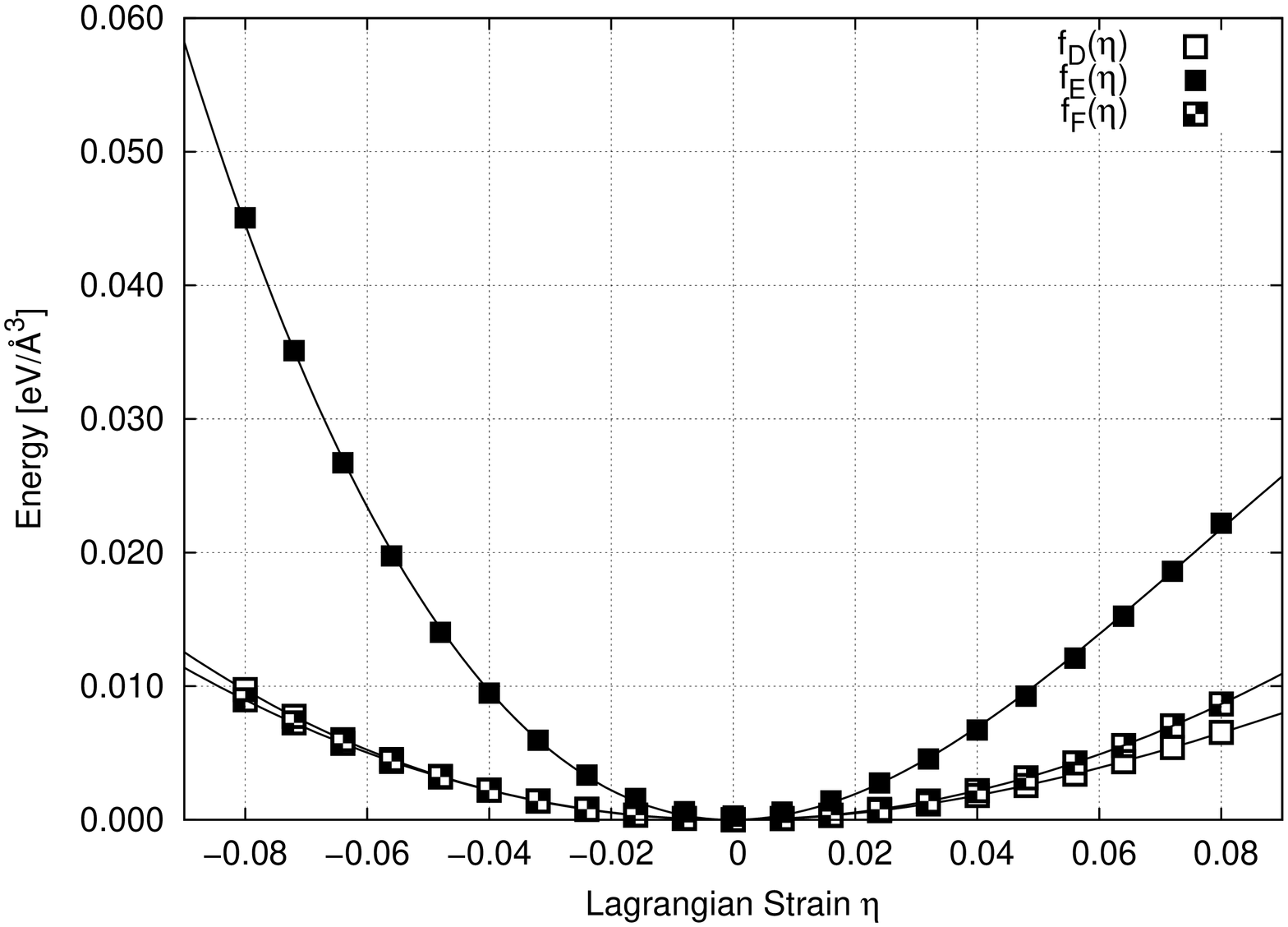}
    \caption{Plots of strain-energy relations for GaN. Squares denote results
             of DFT computations, solid line represents a polynomial 
             fit. See main text for details and Table \ref{tableFormulas} 
             for definitions of $f_A(\eta)$,$f_B(\eta)$,\dots,$f_F(\eta)$.
             \label{figureEnergy}}
\end{figure}

\begin{figure}[!ht]

  \includegraphics[height=0.45\textwidth,angle=270]
        {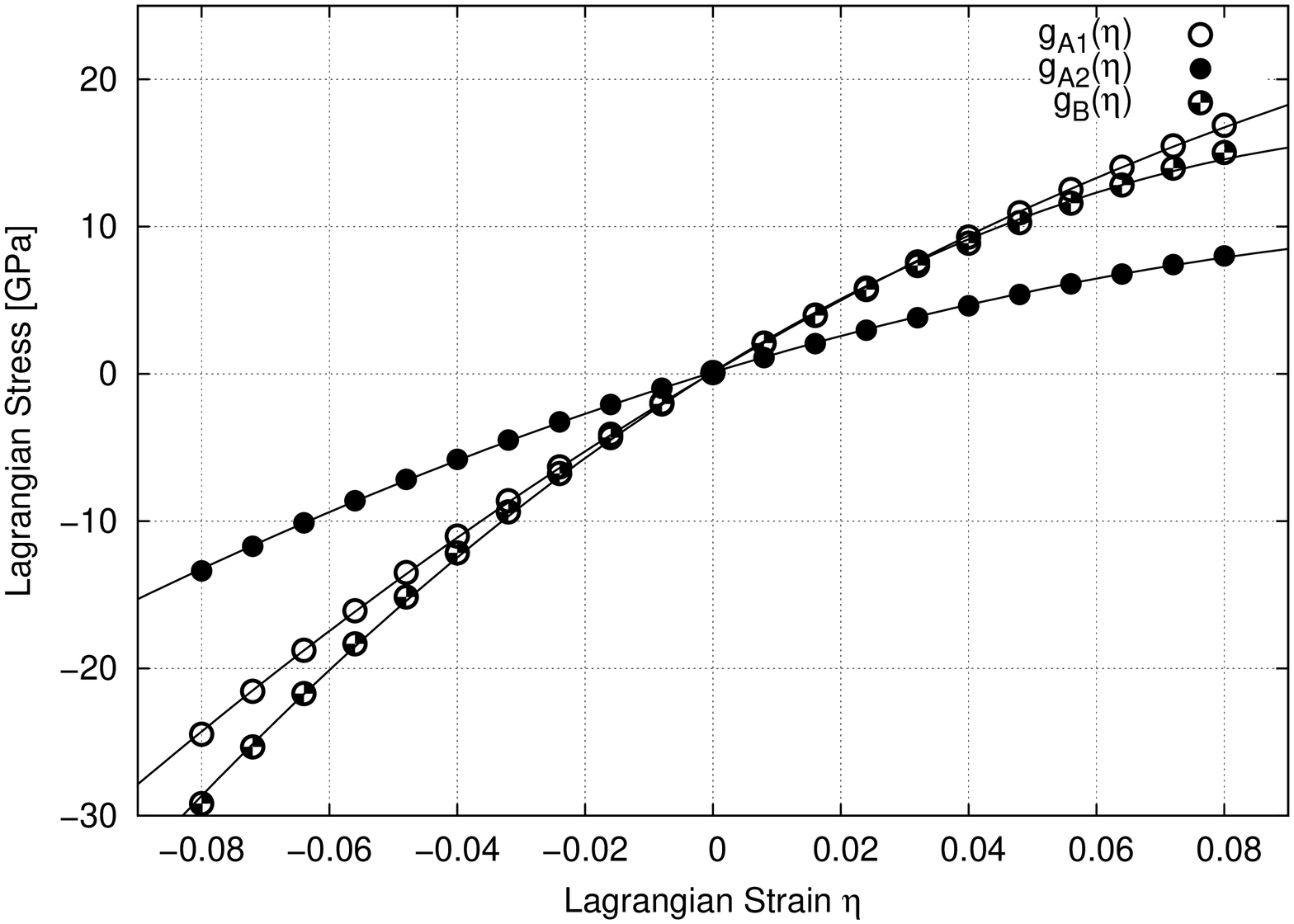}
  \includegraphics[height=0.45\textwidth,angle=270]
        {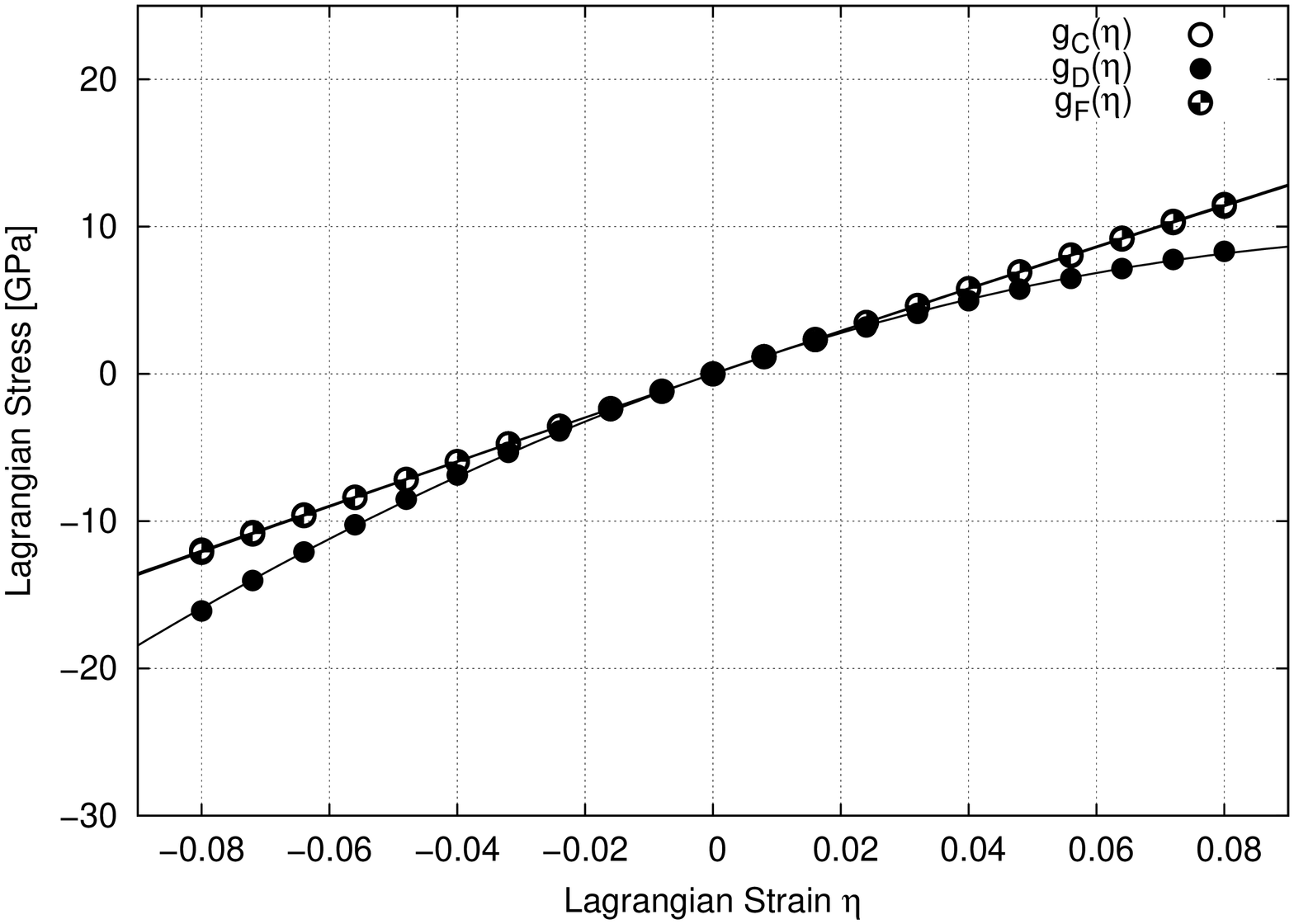}
    \caption{Plots of strain-stress relations for GaN. Circles denote results
             of DFT computations, solid line represents a polynomial 
             fit. See main text for details and Table \ref{tableFormulas} 
             for definitions of $g_{A1}(\eta)$,$g_{A2}(\eta)$,\dots,
             $g_F(\eta)$).
             Curves for $g_{C}(\eta)$ and $g_{F}(\eta)$ coincide because of 
             very similar values of $C_{144}$ and $C_{456}$ in GaN.
             \label{figureStress}        
    }
\end{figure}

Results are presented in Tables \ref{tableConstantsBench} and
\ref{tableConstantsNitrides}. Table \ref{tableConstantsBench} contains our
findings for benchmark materials Si and GaAs, accompanied by available
experimental data and previous theoretical findings within LDA-DFT theory.
Table \ref{tableConstantsNitrides} gives our prediction for the unknown values
of $C_{ijk}$ for cubic nitrides. For completeness, we also provide there our
prediction for second-order elastic moduli and compare them with previous
calculations. \cite{Lepkowski2005} For $c_{ij}$ values, sometimes it was
possible to determine one constant from a few fits [e.g., $c_{44}$ from
coefficients in $f_{C}(\eta)$, $f_{D}(\eta)$, and $f_{F}(\eta)$], obtaining
slightly different results [e.g., for GaN, $c_{44}=145,151,147$ GPa from
$f_{C}(\eta)$, $f_{D}(\eta)$, and $f_{F}(\eta)$ respectively].  In such cases
the average of all obtained values was given in the tables.  The sample plots
of both energy and stress dependencies for GaN together with fitted polynomials
are depicted in Figs.  \ref{figureEnergy} and \ref{figureStress}.    


When analyzing the above results, one has to bear in mind that both measurements
and calculations of the third-order elastic constants are difficult. The
reported experimental results for $C_{ijk}$  are determined with significant
uncertainties and quite often exhibit discrepancies between findings of
different groups (see, e.g., GaAs in Table \ref{tableConstantsBench}).
On the other hand, calculations of subtle third-order effects require
reaching the limits of accuracy of modern quantum codes.

When comparing experimental values with DFT results, it is also worth noticing
that {\it ab initio} calculations are strictly valid for perfect crystalline
structure and in the limit of $0K$ temperature. The experiments, however, are
often performed in conditions far from this idealized case.  Particularly, the
importance of temperature factor can be verified when comparing the results of
measurements for $C_{ijk}$ of Si in temperatures $T=298 K$ and $T=4 K$ given in
Table \ref{tableConstantsBench} (see Ref. \onlinecite{McSkimin1964} for
detailed experimental study). One can observe that for this semiconductor the
values of  constants $C_{144}$ and $C_{123}$ even change their sign, 
when the material is cooled down.

As far as calculations of third-order elastic moduli are concerned, they also
pose a difficult test to {\it ab initio} methods. The determination of
$C_{ijk}$ is sensitive to errors in energy and stress tensor and requires
extremely good convergence of parameters governing the accuracy of
computations, which we believe has been reached in our calculations (see Fig.
1).  The usage of PAW formalism chosen to solve Kohn-Sham equations seems also
not to influence the results significantly, since it has been demonstrated that
properly performed calculations of the static and dynamical properties for broad
range of solids within the PAW, pseudopotential, and LAPW schemes lead to
essentially identical results.\cite{Holzwarth1997} In our opinion, the main
problem lies in the approximations to the exchange-correlation functionals
employed in various calculations. In the present calculations we use GGA-PBE
exchange-correlation functional that is commonly believed to be one of the
best in the market.  However, even for the second-order elastic constants for
GaAs (see Table \ref{tableConstantsBench}), one observes significant
differences between the calculated and measured values.  One possible
origin of these discrepancies might be the commonly known tendency of
calculations based on GGA functional to underestimate binding strength, and
therefore to overestimate lattice constant.   Indeed, our calculations predict
the equilibrium lattice constant of GaAs to be $5.75\; \AA$, considerably
larger than the experimental value of $5.65\; \AA$.\cite{Blakemore1982}  This
is opposite to the local density approximation (LDA), which overestimates
the binding and leads to lattice constants smaller than experimental.

Keeping all the above in mind, we find that the agreement between our
computations and measurements for test cases Si and GaAs is reasonably good
(see Table \ref{tableConstantsBench} for details).  It is also important to
note that values of $C_{ijk}$ calculated both from strain-energy and
strain-stress relations are consistent with each other.  As a cross-check we
additionally verified our approach by calculating  second-order elastic moduli
for GaAs with the aid of the MedeA  package.  \footnote{See
http://www.materialsdesign.com for details about the software.} It uses its
own methodology of calculating $c_{ij}$  on the basis of stress computed by
the VASP code. \cite{LePage2002} We obtained values $c_{11}=99$ GPa, 
$c_{12}=41$ GPa, $c_{44}=51$ GPa, which are in agreement with the results 
given in Table \ref{tableConstantsBench}. 

\begin{figure}[!h]
  \includegraphics[height=0.45\textwidth,angle=270]
        {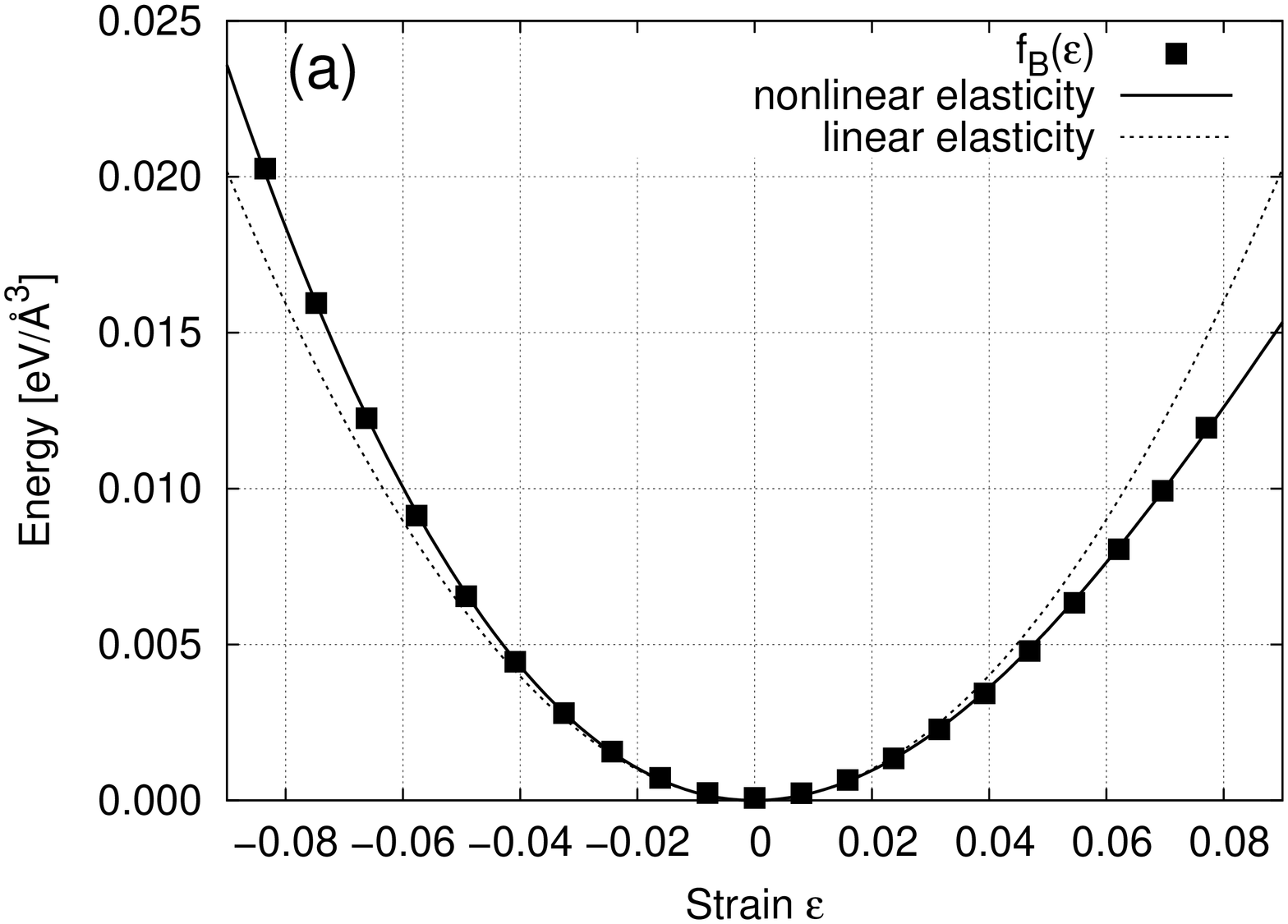}
  \includegraphics[height=0.45\textwidth,angle=270]
        {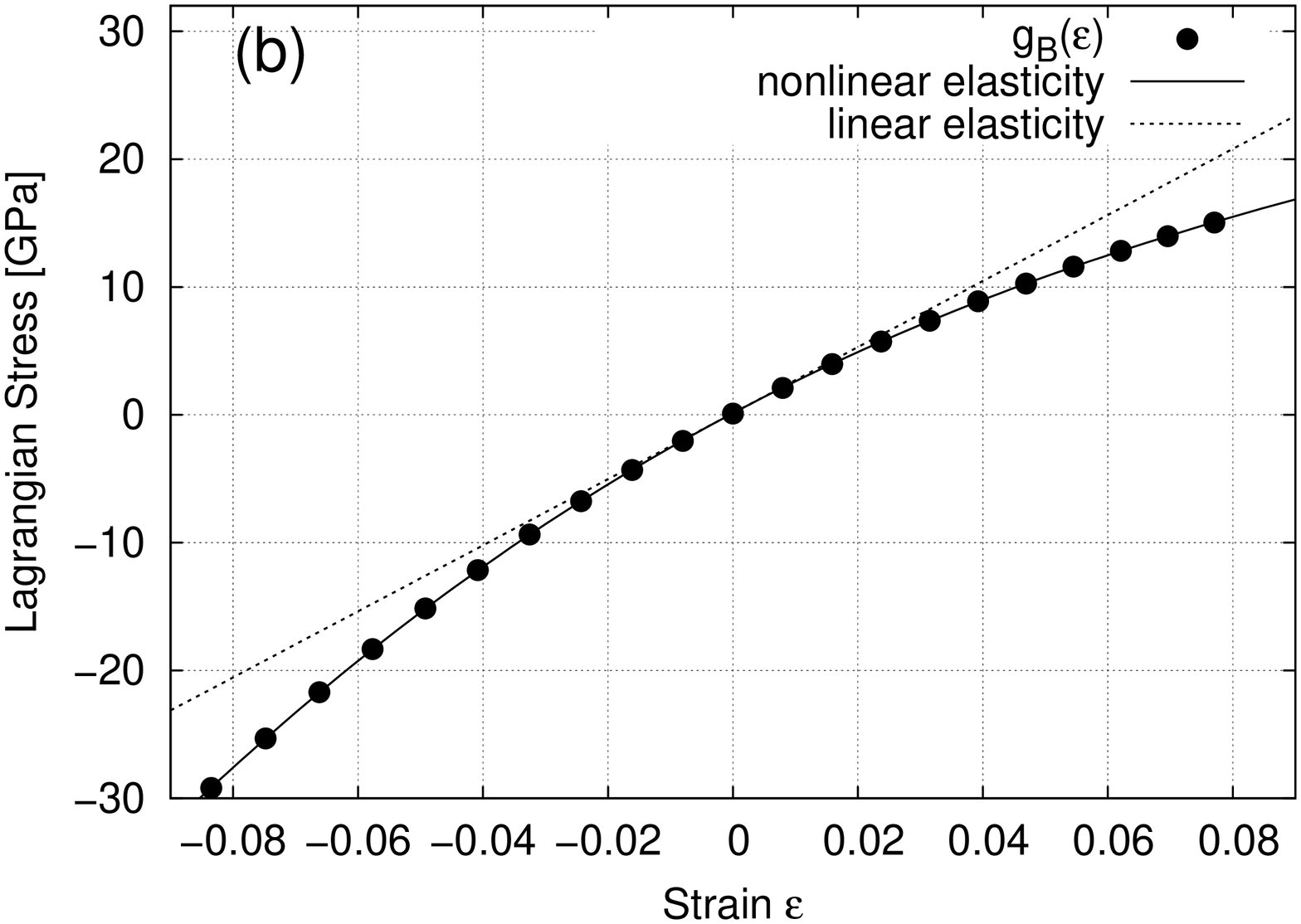}
    \caption{Energy (a) and stress component $t_{3}$ (b) as a function of
             linear strain parameter $\epsilon$ for particular deformation 
             $\bm{\eta}_B$ for GaN cubic crystal.  Full points denote results 
			 of DFT computations, solid and dashed lines indicate the 
             curves obtained from nonlinear and linear elasticity theory 
             respectively.
             \label{figureCompareLinNonlin}        
    }
\end{figure}

Next interesting issue is to examine for which range of deformations the
third-order effects really matter. In Fig. \ref{figureCompareLinNonlin} we
compare energy and stress for the particular deformation $\eta_B$ in GaN
crystal with energy and stress values obtained within linear and nonlinear
elasticity theories. One can clearly see that the linear approach is not
sufficient for strains larger than approximately 2.5\%. It is also worth noting
that for all studied semiconductors and examined range of deformations (i.e.,
with Lagrangian strains up to 8\%) including the terms up to third-order in 
energy expansion [Eq. (\ref{energyStrainRel1})] sufficed to obtain good 
agreement with DFT results.

It is also important to note that a quadratic term in $\bm{\epsilon}$ in the
expression for Lagrangian strain $\bm{\eta}$ [see Eq. (\ref{expansionEta})] is
usually neglected when the second-order elastic constants are determined. For
the third-order elastic constants, such omission is completely unjustified.
For example, the approximation $\bm{\eta} \approx \bm{\epsilon}$ leads to the
following third-order elastic constants for Si, $C_{111}^{\rm{wrong}}=-256\;
\rm{GPa}$, $C_{112}^{\rm{wrong}}=-375\; \rm{GPa}$, $C_{144}^{\rm{wrong}}=94\;
\rm{GPa}$, $C_{155}^{\rm{wrong}}=-130\; \rm{GPa}$, $C_{123}^{\rm{wrong}}=-105\;
\rm{GPa}$, and $C_{456}^{\rm{wrong}}=-2\; \rm{GPa}$, which show significant
disagreement with the results obtained without the aforementioned
simplification (compare results in Table \ref{tableConstantsBench}). As one
would expect, the second-order elastic constants remain virtually unaffected by
the approximation $\bm{\eta} \approx \bm{\epsilon}$, now being $c_{11}=150$
GPa, $c_{12}=62$ GPa, and $c_{44}=73$ GPa.

\begin{table*}
\caption{Comparison of the calculated second- and third-order elastic constants
         for Si and GaAs with the experimental values and previous calculations.
         All data are in GPa. \label{tableConstantsBench}}
\begin{ruledtabular}
\begin{tabular}{l|rrrrrr}
 &  \multicolumn{2}{c}{\hspace{2em}Present results} 
 &  \multicolumn{1}{c}{\hspace{2em}Previous calculations} 
 &  \multicolumn{3}{c}{Experiment}  \\
 
 &  \multicolumn{1}{c}{\hspace{2em}strain-energy} 
 &  \multicolumn{1}{c}{\hspace{2em}strain-stress} & & \\
\hline

\dummy {\bf Si}  & & & \\
$c_{11}$        & 153           	     & 153       
                & 159 \footnotemark[1]       
                & \multicolumn{3}{r}{167 \footnotemark[3]} \\

$c_{12}$        & 65                         & 57   
                & 61 \footnotemark[1]        
                & \multicolumn{3}{r}{65 \footnotemark[3]}  \\
                
$c_{44}$        & 73                         & 75                 
                & 85 \footnotemark[1]        
                & \multicolumn{3}{r}{80 \footnotemark[3]}  \\

$C_{111}$       & -698          & -687
                & -750 \footnotemark[1]      & \hspace{1cm}
                                               -880 \footnotemark[4] & 
                                               -834 \footnotemark[5] &
                                               -825 \footnotemark[6] \\

$C_{112}$       & -451          & -439      
                & -480 \footnotemark[1]      & -515 \footnotemark[4] & 
                                               -531 \footnotemark[5] &
                                               -451 \footnotemark[6]\\
                
$C_{144}$       & 74            &  72      
                &               &  74 \footnotemark[4] &  
                                  -95 \footnotemark[5] &
                                   12 \footnotemark[6] \\

$C_{155}$       & -253          & -252     
                &               & -385 \footnotemark[4] & 
                                  -296 \footnotemark[5] &
                                  -310 \footnotemark[6] \\
                
$C_{123}$       & -112          & -92      
                &  0 \footnotemark[1]        &  27 \footnotemark[4] &  
                                                -2 \footnotemark[5] &
                                               -64 \footnotemark[6] \\
                
$C_{456}$       & -57           & -57 
                & -80 \footnotemark[1]       & -40 \footnotemark[4] &  
                                                -7 \footnotemark[5] &
                                               -64 \footnotemark[6] \\

$C_{144}+2C_{155}$& -430                     & -432
                  & -580 \footnotemark[1]    & -696 \footnotemark[4] 
                                             & -687 \footnotemark[5]
                                             & -608 \footnotemark[6]\\
\dummy {\bf GaAs} & & & \\
$c_{11}$        & 100           & 99     
                & 126 \footnotemark[2]       
                & \multicolumn{3}{r}{113 \footnotemark[7]} \\

$c_{12}$        & 49                         & 41
                & 55 \footnotemark[2]        
                & \multicolumn{3}{r}{57 \footnotemark[7]} \\
                
$c_{44}$        & 52                         & 51               
                & 61 \footnotemark[2]        
                & \multicolumn{3}{r}{60 \footnotemark[7]} \\

$C_{111}$       & -561          & -561 
                & -600 \footnotemark[2]      &  -675 \footnotemark[8] & 
                                                -622 \footnotemark[9] &
                                                -620 \footnotemark[10]      \\

$C_{112}$       & -337          & -318      
                & -401 \footnotemark[2]      &  -402 \footnotemark[8]  &
                                                -387 \footnotemark[9]  &
                                                -392 \footnotemark[10]      \\
                
$C_{144}$       & -14           & -16      
                &  10  \footnotemark[2]      &   -70 \footnotemark[8]  &
                                                   2 \footnotemark[9]  &
                                                   8 \footnotemark[10]      \\
                
$C_{155}$       & -244          & -242      
                & -305 \footnotemark[2]      &  -320 \footnotemark[8]  &
                                                -269 \footnotemark[9]  &
                                                -274 \footnotemark[10]      \\
                
$C_{123}$       & -83           & -70     
                & -94  \footnotemark[2]      &    -4 \footnotemark[8]  &
                                                 -57 \footnotemark[9]  &
                                                 -62 \footnotemark[10]       \\
                
$C_{456}$       & -22           & -22     
                & -43  \footnotemark[2]      &    -69 \footnotemark[8] &
                                                  -39 \footnotemark[9] &
                                                  -43 \footnotemark[10]   \\   
\end{tabular}
\end{ruledtabular}
\footnotetext[1]{ Reference \onlinecite{Nielsen1985}    (LDA).    }
\footnotetext[2]{ Reference \onlinecite{Sorgel1998}     (LDA).    }
\footnotetext[3]{ Reference \onlinecite{McSkimin1953}   ($T=73K$).}
\footnotetext[4]{ Reference \onlinecite{Philip1981}     ($T=4K$). }
\footnotetext[5]{ Reference \onlinecite{Philip1981}    ($T=298K$).}
\footnotetext[6]{ Reference \onlinecite{McSkimin1964}  ($T=298K$).}
\footnotetext[7]{ Reference \onlinecite{Blakemore1982}  
                                        (extrapolation to $T=0K$).}
\footnotetext[8]{ Reference \onlinecite{Drabble1966}   ($T=298K$).}
\footnotetext[9]{ Reference \onlinecite{McSkimin1967}  ($T=298K$).}
\footnotetext[10]{ Reference \onlinecite{Abe1986}       ($T=298K$).}

\end{table*}

\begin{table*}
\caption{ Theoretical predictions for the third-order elastic constants of 
          zinc-blende nitrides - AlN, GaN, and InN.  
          The second-order elastic constants are included
          and compared with previous calculations.
          All data are in GPa.   
          \label{tableConstantsNitrides} }
\begin{ruledtabular}
\begin{tabular}{l|rrr}
 &  \multicolumn{2}{c}{\hspace{2em}Present results} 
 &  \multicolumn{1}{c}{\hspace{2em}Previous calculations} \footnotemark[1] \\ 
 &  \multicolumn{1}{c}{\hspace{2em}strain-energy} 
 &  \multicolumn{1}{c}{\hspace{2em}strain-stress} &  \\
\hline 
\dummy {\bf AlN } &              &              &               \\
$c_{11}$          & 284          & 282          & 267           \\
$c_{12}$          & 167          & 149          & 141           \\
$c_{44}$          & 181          & 179          & 172           \\
$C_{111}$         & -1070        & -1073        &               \\
$C_{112}$         & -1010        & -965         &               \\
$C_{144}$         &    63        &  57          &               \\
$C_{155}$         &  -751        & -757         &               \\
$C_{123}$         &   -78        & -61          &               \\
$C_{456}$         &   -11        & -9           &               \\
\dummy {\bf GaN } &              &              &               \\
$c_{11}$          & 255          & 252          & 252           \\
$c_{12}$          & 147          & 129          & 131           \\ 
$c_{44}$          & 148          & 147          & 146           \\
$C_{111}$         & -1209        & -1213        &               \\
$C_{112}$         & -905         & -867         &               \\
$C_{144}$         & -45          & -46          &               \\
$C_{155}$         & -603         & -606         &               \\
$C_{123}$         & -294         & -253         &               \\
$C_{456}$         & -48          & -49          &               \\

\dummy {\bf InN}  &               &             &               \\
$c_{11}$          & 160           & 159         & 149           \\
$c_{12}$          & 115           & 102         &  94           \\
$c_{44}$          & 78            & 78          &  77           \\
$C_{111}$         & -752          & -756        &               \\   
$C_{112}$         & -661          & -636        &               \\
$C_{144}$         & 16            & 13          &               \\
$C_{155}$         & -268          & -271        &               \\
$C_{123}$         & -357          & -310        &               \\
$C_{456}$         & 14            & 15          &               \\ 
\end{tabular}
\end{ruledtabular}

\footnotetext[1]{ Reference \onlinecite{Lepkowski2005}  (GGA).    }
\end{table*}

\section{ Relation to pressure dependent elastic constants 
          \label{PressureConstants} }
In the case of materials under large hydrostatic pressure it is useful to
describe the nonlinear elastic properties using the concept of pressure
dependent elastic constants $c_{ij}(P)$.  For many applications, it is
sufficient to consider only terms linear in the external hydrostatic pressure
\begin{eqnarray}
c_{11}^P(P) &\approx& c_{11} + {c'}_{11} P,  \nonumber \\
c_{12}^P(P) &\approx& c_{12} + {c'}_{12} P,  \\
c_{44}^P(P) &\approx& c_{44} + {c'}_{44} P, \nonumber
\end{eqnarray}
with pressure derivatives ${c'}_{ij}$ being material parameters.
Naturally, the information about ${c'}_{ij}$ can be recovered from third-order
elastic constants.  The necessary formulas are given below \cite{Birch1947}
\begin{eqnarray}
\label{pressureDer}
{c'}_{11}&=&-\frac{2 C_{112} + C_{111} + 2 c_{12} + 2 c_{11}}{2c_{12} + c_{11}},
                   \nonumber \\
{c'}_{12}&=&-\frac{C_{123} + 2 C_{112} - c_{12} -c_{11}}{2c_{12} + c_{11}}, \\
{c'}_{44}&=&-\frac{2 C_{155} + C_{144} + c_{44} + 2 c_{12} + c_{11} }
                { 2c_{12} + c_{11}}.
            \nonumber                           
\end{eqnarray}
Results for pressure derivatives ${c'}_{ij}$ calculated on the basis of our
estimates for second- and third-order elastic constants are shown in Tables
\ref{tableDerivativesGaAs} and \ref{tableDerivativesNitrides}. 

Table \ref{tableDerivativesGaAs} provides comparison with experimental
results  for Si \cite{Beattie1970} and GaAs.\cite{McSkimin1967a} The agreement
is very good and shows that the results from the strain-energy relation
reproduce the experimental values slightly better than findings based on
strain-stress formula.  

Table \ref{tableDerivativesNitrides} contains values of ${c'}_{ij}$ for
zinc-blende nitrides and compares the present calculation with our previous
work. \cite{Lepkowski2005}  In Ref. \onlinecite{Lepkowski2005}, the following
approach for the determination of pressure dependence of the second-order
elastic constants has been used. First, the hydrostatic strain (corresponding
to the external pressure $P$) has been applied to a crystal, and then the
crystal has been additionally deformed to determine the pressure dependent
elastic constants.  The DFT results for the total elastic energy combined with
the strain-energy relation have enabled us to determine $c_{ij}(P)$ as well as
${c'}_{ij}$. We would like to stress that the additional noninfinitesimal
strain has not always been trace-free just leading to a spurious hydrostatic
component that has modified external hydrostatic pressure.  Therefore, we
believe that the approach employed in the present paper is not only more
direct, but also slightly more accurate.  The discrepancies between our present
and previous results can also be partly ascribed to the methodological
differences, such as different exchange-correlation functional used and
slightly different calculation parameters (Brillouin zone sampling, energy
cutoffs etc.). 
 
\begin{table}[!ht]
\caption{Pressure derivatives of second-order elastic constants for Si 
         and GaAs calculated on the basis of Eqs. (\ref{pressureDer}) . 
         For comparison experimental findings are included. 
         \label{tableDerivativesGaAs}}
\begin{ruledtabular}
\begin{tabular}{l|ccc}
                                  &     \multicolumn{2}{c}{Present results} 
                                  & Experiment  \\
                  & strain-energy & strain-stress 
                                  &  \\
\hline 
\dummy {\bf Si} &   & &  \\
${c'}_{11}$  &  4.09   &  4.30   &  4.19 \footnotemark[1] \\
${c'}_{12}$  &  4.34   &  4.43   &  4.02 \footnotemark[1] \\
${c'}_{44}$  &  0.27   &  0.34   &  0.80 \footnotemark[1] \\ 
\dummy {\bf GaAs}    & &         &       \\
${c'}_{11}$  &  4.71   &  5.06   &  4.63 \footnotemark[2] \\
${c'}_{12}$  &  4.56   &  4.67   &  4.42 \footnotemark[2] \\
${c'}_{44}$  &  1.27   &  1.48   &  1.10 \footnotemark[2] 
\end{tabular}
\end{ruledtabular}
\footnotetext[1]{ Reference \onlinecite{Beattie1970}   (T=4K).  }
\footnotetext[2]{ Reference \onlinecite{McSkimin1967a} (T=298K).}
\end{table}

\begin{table}[!ht]
\caption{Prediction of pressure derivatives of second-order elastic constants  
         for zinc-blende nitrides AlN, GaN, and InN calculated on the basis of 
         Eqs. (\ref{pressureDer}). For comparison, results of previous
         calculations employing different methodology are included. 
         \label{tableDerivativesNitrides}}
\begin{ruledtabular}
\begin{tabular}{l|ccc}
                                  &     \multicolumn{2}{c}{Present results} 
                                  & Previous calculations \footnotemark[1] \\
                  & strain-energy & strain-stress 
                                  &  \\
\hline 
 \dummy {\bf AlN}    &   &   &  \\
 ${c'}_{11}$  &  3.53   &  3.68   &  5.21  \\
 ${c'}_{12}$  &  4.12   &  4.17   &  4.26  \\
 ${c'}_{44}$  &  1.03   &  1.20   &  1.69  \\
   \dummy {\bf GaN}    &   &   &  \\
 ${c'}_{11}$  &  4.03   &  4.28   &  4.17  \\
 ${c'}_{12}$  &  4.56   &  4.64   &  3.50  \\
 ${c'}_{44}$  &  1.01   &  1.18   &  1.12  \\
   \dummy {\bf InN}    &   &   &  \\
 ${c'}_{11}$  &  3.89   &  4.15   &  4.58  \\
 ${c'}_{12}$  &  5.00   &  5.08   &  4.37  \\
 ${c'}_{44}$  &  0.13   &  0.24   &  0.66 
\end{tabular}
\end{ruledtabular}
\footnotetext[1]{Reference \onlinecite{Lepkowski2005}.  }
\end{table}

\section{ Relation to Gr\"uneisen constants of long-wavelength
 acoustic modes \label{GrueneisenConstants}}

The mode Gr\"uneisen constants constitute a group of important coefficients,
which characterize anharmonic properties of crystals. These quantities are
frequently encountered in theory of phonons and in the description of
thermodynamical properties of solids. The mode Gr\"uneisen constants are
defined as follows:
\begin{equation}
 \gamma(\mathbf{q},\mathbf{j})= - \frac{ \partial \ln 
                                  \omega(\mathbf{q}, \mathbf{j}) }
                             { \partial \ln   V }
                      = - \frac{V}{ \omega(\mathbf{q},\mathbf{j}) } 
                          \frac{ \partial \omega(\mathbf{q},\mathbf{j}) }
                             { \partial V},
\end{equation}
where $\omega$ denotes the frequency of phonon with wave vector $\bf{q}$
and polarization vector $\mathbf{j}$. $V$ stands here for volume of the 
crystal.

On the basis of continuum limit, one may express mode Gr\"uneisen constants for
long-wavelength acoustic modes in terms of second- and third-order elastic
constants.  The necessary expressions used here have been  given by Mayer and
Wehner. \cite{Mayer1984}  The results for $\gamma({\bf q},{\bf j})$ obtained
from our strain-energy estimates of elastic moduli are given in Table
\ref{tableGrueneisen}. 

Comparison with the experimental data available for Si shows that results
calculated by us often differ significantly from experimental findings.  The
discrepancy is particularly pronounced for transverse modes (i.e.,
$\gamma((\epsilon,0,0),\rm{TA})= \gamma((\epsilon,\epsilon,0),\rm{TA}_{z})$
and $\gamma((\epsilon,\epsilon,0),\rm{TA}_{xy})$) for which the magnitudes
of Gr\"uneisen constants are much smaller than for longitudinal modes.  In our
opinion, this indicates that $\gamma({\bf q},{\bf  j})$ are quite sensitive to
inaccuracies in $C_{ijk}$ values.  Therefore, one has to treat our prediction
for mode Gr\"uneisen constants in zinc-blende nitrides rather as a quite crude
approximation.  Nevertheless, it could be an interesting subject of further
studies to  compare the above results with {\it ab initio} phonon calculations
performed via density functional perturbation theory. More detailed
experimental studies for a broader range of materials could also shed more 
light on the value of the presented theoretical predictions.

\begin{table}
 \caption{Gr\"uneisen constants $\gamma({\bf q}, {\bf j})$ for long-wavelength
 acoustic modes.  Experimental results for Si were given in Ref.
 \onlinecite{Mayer1984} on the basis of ultrasound measurements data from Ref.
 \onlinecite{McSkimin1964}.  Theoretical prediction for 
 $\gamma({\bf q}, {\bf j})$ were calculated on the basis of strain-energy 
 values for $c_{ij}$ and $C_{ijk}$.
 \label{tableGrueneisen} }
 \begin{ruledtabular}
 \begin{tabular}{l|cc|ccc}
  & Experiment & Theory & 
    \multicolumn{3}{c}{Theory} \\
  & Si   
           &   Si  &   AlN   &   GaN   &   InN  \\
 \hline
 \dummy ${\bf q}=(\epsilon, 0, 0)$  &   &   &   &   &  \\
 $\gamma(\rm{LA})     $  &  1.108  &  1.098  &  1.115  &  1.279  &  1.415 \\
 $\gamma(\rm{TA})     $  & 0.324  &  0.006  &  0.423  &  0.459  & -0.055 \\
 \dummy ${\bf q}=(\epsilon, \epsilon, 0)$  &   &   &   &   &  \\
 $\gamma(\rm{LA})$       & 1.109  &  0.999  &  1.066  &  1.226  &  1.218 \\
 $\gamma(\rm{TA_{xy}})$  & -0.049  & -0.301  & -0.684  & -0.613  & -1.771 \\
 $\gamma(\rm{TA_{z}}) $  &  0.324  &  0.006  &  0.423  &  0.459  & -0.055 \\
 \dummy ${\bf q}=(\epsilon, \epsilon, \epsilon)$  &   &   &   &   &  \\
 $\gamma(\rm{LA})     $  &  1.081  &  0.973  &  1.056  &  1.214  &  1.173 \\
\end{tabular}
\end{ruledtabular}
\end{table}

\section{ Conclusions \label{Conclusions} }
We have presented a detailed {\it ab initio} study of third-order elastic constants
$C_{ijk}$ for selected semiconductors - Si, GaAs, and zinc-blende nitrides AlN,
GaN, and InN.  Even though third-order effects  are very subtle,  we showed
that it is possible to estimate them by means of density functional theory on
the GGA level. We have used two approaches involving either strain-energy or
strain-stress relations, obtaining consistent results from both of them.  To
benchmark the reliability of the presented method, we have compared our theoretical
results for Si and GaAs with available experimental findings. The agreement is
reasonable, however, particularly for moduli of smaller magnitude (e.g., for
examined cases typically $C_{144}$ and $C_{456}$)  relative differences are
significant. In our opinion, they can be ascribed to three main factors:
shortcomings of GGA-DFT theory, lack of temperature effects in our calculations
(experimental results for $C_{ijk}$ are usually obtained in room temperature),
and measurement uncertainties. We have also underlined the relation of
third-order elastic constants to other anharmonic properties. On the basis of
the {\it ab initio} results for $C_{ijk}$, we have computed the pressure derivatives
of second-order elastic moduli and provided rough estimations for Gr\"uneisen
constants of long-wavelength acoustic modes. We believe that DFT estimates of
third-order elastic constants can be a very useful tool in modeling
semiconducting nanostructures, in which nonlinear effects often play an
important role.

\begin{acknowledgments}
This work was partly supported by the Polish State Committee for Scientific
Research, Project No. 1P03B03729.
One of the authors (M.\L{}.) wishes to acknowledge useful discussion 
with Alexander Mavromaras from Materials Design.
\end{acknowledgments}

\bibliography{bibliography}
\end{document}